\documentclass[twocolumn,showpacs,preprintnumbers,amsmath,amssymb]{revtex4}

\usepackage{graphicx}
\usepackage{dcolumn}
\usepackage{bm}
\usepackage{subfigure}

\begin{document}

\title{Unfolding Substructures of Complex Networks by Coupling Chaotic Oscillators beyond Global Synchronization Regime}
\author{Zhao Zhuo$^{1}$}
\email{zhzh7532@mail.ustc.edu.cn}
\author{Shi-Min Cai$^{1}$}
\email{csm1981@mail.ustc.edu.cn}
\author{Jie Zhang$^{2}$}
\email{jzhang080@gmail.com}
\author{Zhong-Qian Fu$^{1}$}
\email{zqfu@ustc.edu.cn}

\affiliation{$^{1}$Department of Electronic Science and
Technology, University of Science and Technology of China, Hefei
Anhui, 230026, PR China \\
$^{2}$Centre for Computational Systems Biology, Fudan
University, Shanghai 200433, PR China}

\date{\today}

\begin{abstract}
In the past decade, synchronization on complex networks has
attracted increasing attentions from various research disciplines.
Most previous works, however, focus only on the dynamic behaviors of
synchronization process in the stable region, i.e., global synchronization. In this letter, we
demonstrate that synchronization process on complex networks
can efficiently reveal the substructures of networks when
the coupling strength of chaotic oscillators is under the lower boundary of stable region.
Both analytic and numerical results
show that the nodes belonging to the same component in
the hierarchical network are tightly clustered according to the Euclidean
distances between the state vectors of the corresponding oscillators, and different levels
of hierarchy can be systematically unfolded by gradually tuning the coupling
strength. When the coupling strengths exceed the upper boundary of
stable region, the hierarchy of the network cannot be recognized
by this approach. Extensive simulations suggest that our method may
provide a powerful tool to detect the
hierarchical community structure of complex systems and deep
insight into the relationship between structure and dynamics of complex systems.

\end{abstract}


\maketitle

Synchronization phenomena of interacting units, such as clapping
peoples, fireflies and pendulums, have been noticed by scientists
for a long time \cite{Hugenii1673,Pecora1990,Neda2000,Pikovsky2002}.
For convenience of research, the interacting relations between
units are usually abstracted as a network, whose nodes represent units and
edges indicate interactions between them. With the
dramatic progress made in complex network science in the last decade,
researchers from various disciplines such as biology, physics and engineering communities have
begun to explore a common interesting problem: the relation between
topology of complex interactions and the emergent synchronization process
\cite{Strogatz2001,Oh2005,Zhou2006,Arenas2007,Arenas2008,Yu2009,Nishikawa2010,Zhuo2011,Zhang2009,Zhang2010}.
On one hand, lots of works have investigated how synchronization
process appears under different topology. For instance, the stability,
one of the most important properties of synchronization, has been
studied by the master stable function (MSF) method when the
coupling networks have different topological properties like degree
distribution, average path length, cluster coefficient and edge
weight \cite{Pecora1998,Barahona2002,Nishikawa2003,Hong2004,Li2004,Belykh2004,Yook2006}.
In most cases, the stability is proportional to the eigenratio
$\lambda_{max}/\lambda_2$, where $\lambda_{max}$ is the largest
eigenvalue of Laplacian matrix of network and $\lambda_2$ is the
second smallest one. Based on the studies of stability, many methods
are introduced to enhance the synchronizability via edge
betweenness, topology modification, optimization, adaptive
evolution, and so on
\cite{Chavez2005,Zhao2005,Yin2006,Donetti2005,Nishikawa2006a,Nishikawa2006b,zhou2006,Yan2009}.
On the other hand, few works focus on how topology of complex interactions can be reflected
by synchronization process. In \cite{Arenas2006}, the hierarchical
structure of network is gradually revealed from
phase correlations in Kuramoto model \cite{Kuramoto1984} when the system evolves
to the stable state. Recently, Ren \emph{et.al} developed a
universal approach to predict the exact network topology based solely
on measuring the dynamical correlations of time series generated by the
global synchronization \cite{Ren2010}.

In this letter, we focus on the problem of how substructures
of complex networks can be reflected in the unstable synchronization
processes taking place on networks. Both the analytical and
numerical results show that if the coupling strength is under the
lower boundary of stable region, the nodes belonging to different
components of hierarchical structure can be directly
distinguished by the state of the corresponding oscillator,
which dramatically reduces the computational complexity of
obtained cross-correlation among multiple time series in traditional
methods. Moreover, for a hierarchical network, we can
unravel different levels of hierarchies in a
top-down (or bottom-up) way by decreasing (or increasing) coupling
strength gradually. However, if the coupling strength exceeds the
upper boundary of the stable region, we cannot find the cluster
structures of nodes even they couple in a hierarchical topology. In
the following, we will first show that the states of coupling
oscillators and substructure of network topology are both associated
with the Laplacian eigenvalues and eigenvectors in a theoretical
way.

Consider a generic system composed of $N$ coupling chaotic oscillators,
each one is represented by a $m$-dimensional state vector
$\mathbf{x}_i$ and ruled by the differential equations
$\mathbf{\dot{x}}_i = \mathbf{F}(\mathbf{x}_i)$. The network
topology are defined by the Laplacian matrix $\mathbf{L} =
(l_{ij})_{N\times N}$, in which $l_{ij}=1$ if oscillator $i$ and $j$ are coupled, $l_{ij} =0$
otherwise, and $l_{ii}=k_i$ (the degree of oscillator $i$). Therefore,
we have the evolution equation of system
\begin{equation}
\mathbf{\dot{x}}_i=\mathbf{F}(\mathbf{x}_i)-\sigma\sum_{j=1}^{N}l_{ij}\mathbf{H}(\mathbf{x}_j).\label{evolving}
\end{equation}
Here we only consider undirected and binary networks,

As the Laplacian matrix has zero row-sum, there is a global
synchronization state of oscillators,
\begin{equation}
\mathbf{x}_1(t)\rightarrow
\mathbf{x}_2(t)\rightarrow...\rightarrow\mathbf{x}_N(t)\rightarrow\mathbf{s}(t).\label{synch}
\end{equation}
Let $\delta\mathbf{x}_i = \mathbf{x}_i-\mathbf{s}$ be the deviation
of the \emph{i}th oscillator from the synchronization manifold, we
have
\begin{equation}
\delta\mathbf{\dot{x}}_i=D\mathbf{F}(\mathbf{s})\delta\mathbf{x}_i
-\sigma
D\mathbf{H}(\mathbf{s})\sum_{j=1}^{N}G_{ij}\delta\mathbf{x}_j.\label{variation}
\end{equation}
Let $\lambda_1<\lambda_2<...<\lambda_N$ denote Laplacian eigenvalues
and $\mathbf{v}_1, \mathbf{v}_2,...,\mathbf{v}_N$ denote the
associated Laplacian eigenvectors. To diagonalize
the variational equations, we rewrite $\delta\mathbf{x}$ as
$\delta\mathbf{x} = \mathbf{O}\mathbf{\xi}$, where $\mathbf{O} =
[\mathbf{v}_1, \mathbf{v}_2,...,\mathbf{v}_N]$ and
$\mathbf{\xi}=[\xi_1, \xi_2,...,\xi_N]^T$. Hence, we obtain the $N$
decoupled eigenmodes of variational equations
\begin{equation}
\mathbf{\dot{\xi}}_i=[D\mathbf{F}(\mathbf{s})-\sigma\lambda_li
D\mathbf{H}(\mathbf{s})]\mathbf{{\xi}}_i, i = 1, 2, ... , N,
\label{MSF}
\end{equation}
Eq. (\ref{MSF}) is called the MSF of the complex system. $\xi_i$ is
the coefficient corresponding to eigenvector $\mathbf{v}_i$ in the
linear combination. The eigenvector $\mathbf{v}_1=[1,1,...,1]$ of
$\lambda_1=0$ corresponds to the synchronization manifold. The
global synchronization of oscillators is achieved if
$\sigma\lambda_i (i\geq2)$ totally falls in the stable region
$\mathbf{S}$ which makes the maximum transverse Lyapunov exponents
of Eq. (\ref{MSF}) ($i\geq2$) be negative. The stable region
$\mathbf{S}$ can be empty($\phi$), bounded($[\alpha_1\ \alpha_2]$)
or unbounded($[\alpha_1\ \infty]$) with different chaotic
oscillators and coupling function $\mathbf{H}$.

To identify different communities in a hierarchical network, we
set coupling strength out of the stable region to ensure the
differences between oscillator states remain. It's discovered that
eigenvalues and eigenvectors of Laplacian matrix are strongly
associated to substructures of network \cite{Capocci2004,huang2006}.
In a network with $m$ communities, the former $m-1$
non-trivial eigenvalues (from $\lambda_2$ to $\lambda_m$) are
approximately equal to 0 and much smaller than the others, which
leads to a gap between the former $m$ eigenvalues and the
others. And if a network is organized in a hierarchical way, there
are more gaps between eigenvalues, which implies the different
levels of hierarchy \cite{Arenas2006}. The eigenvectors of the
former $m-1$ eigenvalues describe the organization of network.
Elements of eigenvectors corresponding to nodes in the same
community share approximately the same value. In the hierarchical
network, the eigenvectors corresponding to the eigenvalues before the
first gap only show a coarse macroscopic organization of the highest
hierarchy. When eigenvalues increasing and crossing following gaps,
their eigenvectors can suggest a finer mesoscopic organization of
lower hierarchies. We choose the coupling strength according
to the inequality $\sigma\lambda_m<\alpha_1<\sigma\lambda_{m+1}$,
which makes the coefficients $\xi_2$ to $\xi_m$ diverge so that the
final states of oscillators are linear combinations of eigenvectors $\mathbf{v}_2$
to $\mathbf{v}_m$ and the perturbations of $\mathbf{v}_{m+1}$ to
$\mathbf{v}_N$ vanish for sufficient evolving time. By increasing
$m$, the more eigenvectors corresponding to larger eigenvalues
are added as the basis of the linear combination
of state vectors and more details of hierarchical
structure are uncovered by the state vectors.

\begin{figure}[!t]
\centering {\includegraphics[width=3.5in]{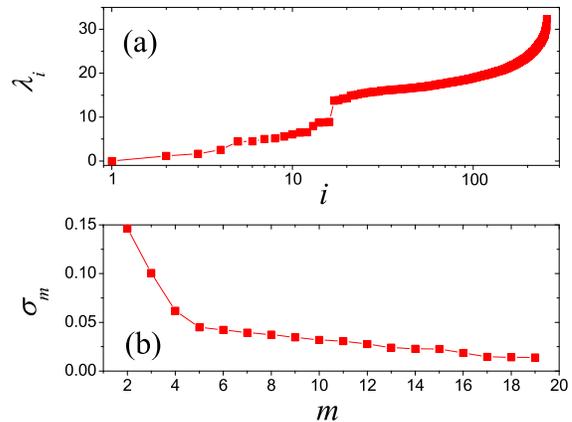}} \caption{
(a)eigenvalues $\lambda_i$ of Laplacian matrix of
the hierarchical network considered.(b) Coupling strength
$\sigma_m$ versus $m$.}\label{fig:lc}
\end{figure}

We simulate the synchronization process of coupling chaotic
oscillators on a hierarchical network to confirm the above
theoretical analysis. We firstly choose the R\"ossler oscillators as
an example of our investigation, whose stable region is bounded
in [$\alpha_1\ \alpha_2$]. Their state vectors and
differential equations can be written as $\mathbf{x}=[x,y,z]^T$ and
$\mathbf{F}=[-(y+z),x+0.2y,0.2+z(x-9)]^T$. Units are coupled by the
linear coupling function $\mathbf{H(x)}=[x,0,0]^T$. We therefore can
get $\alpha_1\approx0.2$, and $\alpha_2\approx5$ according to Eq.
\ref{MSF}. The hierarchical network is firstly proposed in
\cite{Arenas2006}, which includes two hierarchical levels of
components. Specifically, it consists 256 nodes, in which 16
non-overlapping clusters each containing 16 nodes represent the
first hierarchical level and 4 larger non-overlapping clusters each
containing four ones from the first level consist the second
hierarchical level. The connections of nodes at the first level
($z_{in1}$), the second level ($z_{in2}$) and with the rest nodes
($z_{out}$) satisfy $z_{in1}+z_{in2}+z_{out}=18$, and the
hierarchical network is therefore indicated as $z_{in1}-z_{in2}$.

\begin{figure*}[!t]
\centering {\includegraphics[width=7in]{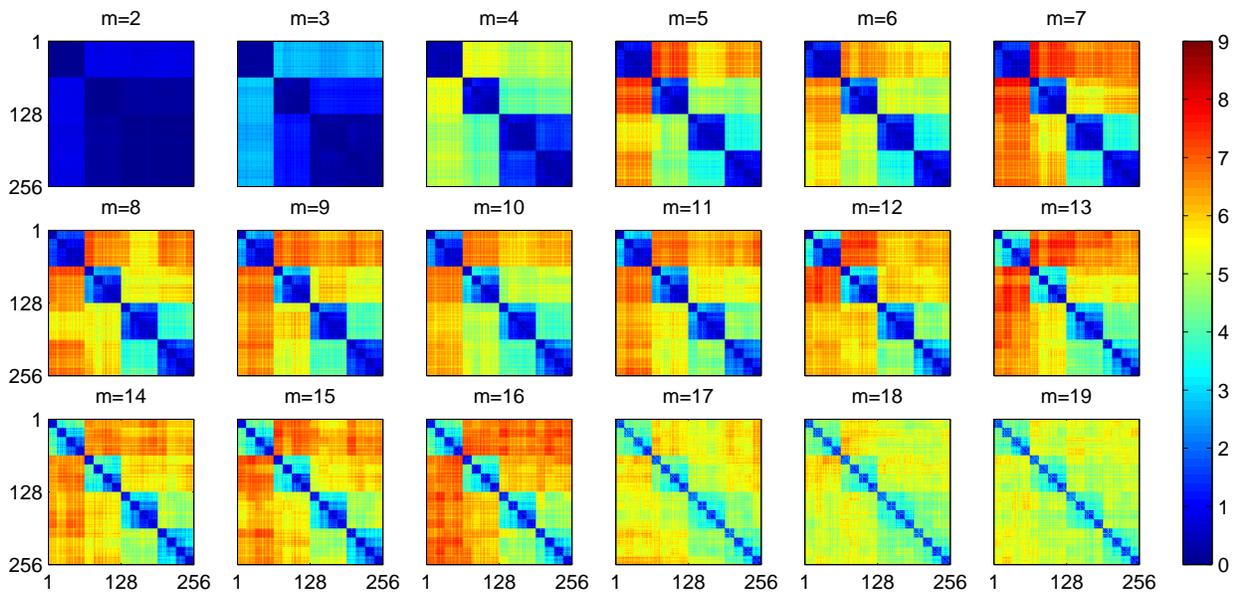}} \hspace{0.3in}
\caption{(Color online) Distance matrix of 14-3 network with $m$
changing from 2 to 19. With small $m$, only large clusters are recognized.
Details of hierarchical structure emerge when the coupling strength decreases
(i.e., $m$ increases).}\label{fig:dis}
\end{figure*}

The synchronization process is performed on the $14-3$ hierarchical
network. The coupling strength is defined as
\begin{equation}
\sigma_m=\frac{1}{2}[\frac{\alpha_1}{\lambda_m}+\frac{\alpha_1}{\lambda_{m+1}}],m=2,3,...,N-1.\label{coupling},
\end{equation}
which guarantees $\sigma\lambda_m<\alpha_1<\sigma\lambda_{m+1}$
(i.e., the global synchronization is impossible). We show the
Laplacian eigenvalues of the hierarchical network in Fig.
\ref{fig:lc}(a), in which there are two obvious gaps that
imply the two levels of hierarchy. As shown in Fig.
\ref{fig:lc}(a), the first gap is between $\lambda_4$ and
$\lambda_5$, while the second one is between $\lambda_{16}$ and
$\lambda_{17}$. Thus, we are able to unfold the hierarchical
structure gradually as $m$ increases from 2 to 19. By considering
the eigenvalues and lower limit of the stable region of R\"ossler
oscillator, we set the coupling strength according to Eq.
\ref{coupling}, which monotonously decreases with $m$ (see in Fig.
\ref{fig:lc}(b)). Moreover, the maximum coupling strength $\sigma_2$
is also constrained by $\sigma_2\lambda_N<\alpha_2$ to make the
coefficient $\xi_i$ corresponding to $\lambda_i (i\geq m+1)$
converge and perturbations of $\mathbf{v}_{m+1}$ to $\mathbf{v}_N$
vanish for sufficient evolving time.

We use the Euler method with time step $\Delta t=0.001$ and total
evolving time $T=100$ to numerically simulate the synchronization
process. The initial states of R\"ossler oscillators (i.e., the
nodes of hierarchical network) are uniformly distributed in the
interval $[0,1]$. The difference of state between the nodes is
suggested by their Euclidean distance $d_{ij} =
\sqrt{(x_i-x_j)^2+(y_i-y_j)^2+(z_i-z_j)^2}$. Figure. \ref{fig:dis}
shows the distance matrices at $T=100$ at a changing coupling
strength $\sigma_m$ from $m=2$ to $m=19$. Each distance matrix is
averaged over a hundred runs of synchronization process. We find
that the blocks consisting of trivial values exist in all distance
matrices. Inside each block, the distances between nodes are small,
while the distance between nodes across different blocks is large
enough to unfold the hierarchy of network. Specifically, with
smaller $m$, the coupling strength is large, which causes
only the coefficients $\xi_i$ corresponding to the smallest
eigenvalues to diverge. In these cases, we can only recognize the
larger clusters at the second hierarchical level. When $m \geq 5$,
the coefficients corresponding to the eigenvalues after the first gap
begin to diverge, which suggests that in the distance matrices, the
4 non-overlapping components at the second hierarchical level split
into smaller components at the first hierarchical level. As $m$
increases further from 5 to 16, the eigenvectors
corresponding to the larger eigenvalues can provide the
difference between nodes that belong to the same component at the
second hierarchical level, as is manifested in Fig \ref{fig:dis}. Thus, the 16
non-overlapping components at the second hierarchical level become
much more obvious and the two-level hierarchy of network is
gradually revealed. For $m \geq 5$, the coupling strength decrease
to trivial values so that the synchronization process is
weakened, which can help reveal more detailed structure of
hierarchical network.

\begin{figure}[!t]
\centering
\subfigure{\label{fig:hhs}\includegraphics[width=3.3in]{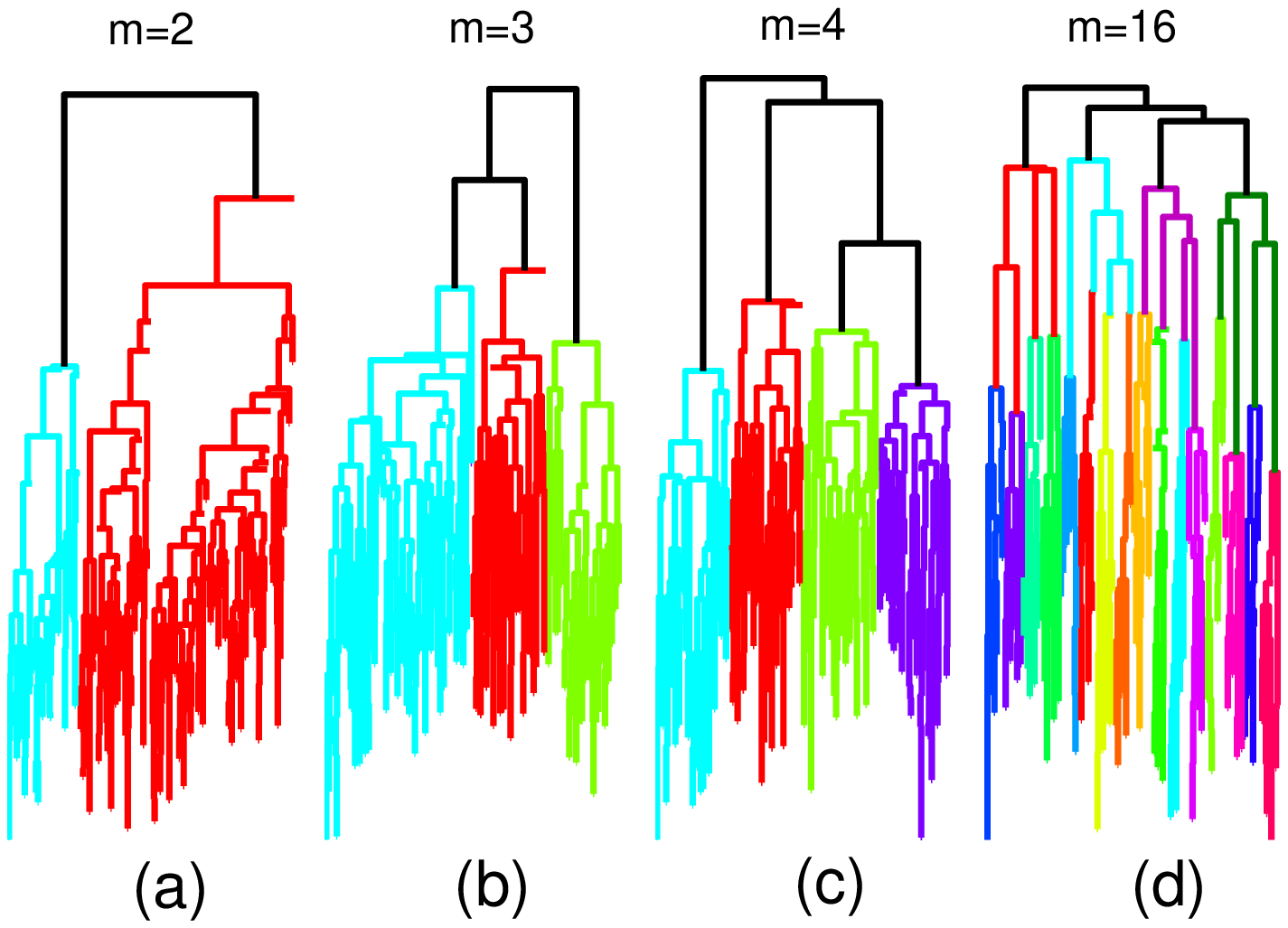}}
\hspace{0.3in}
\\
\subfigure{\label{fig:net}\includegraphics[width=3.3in]{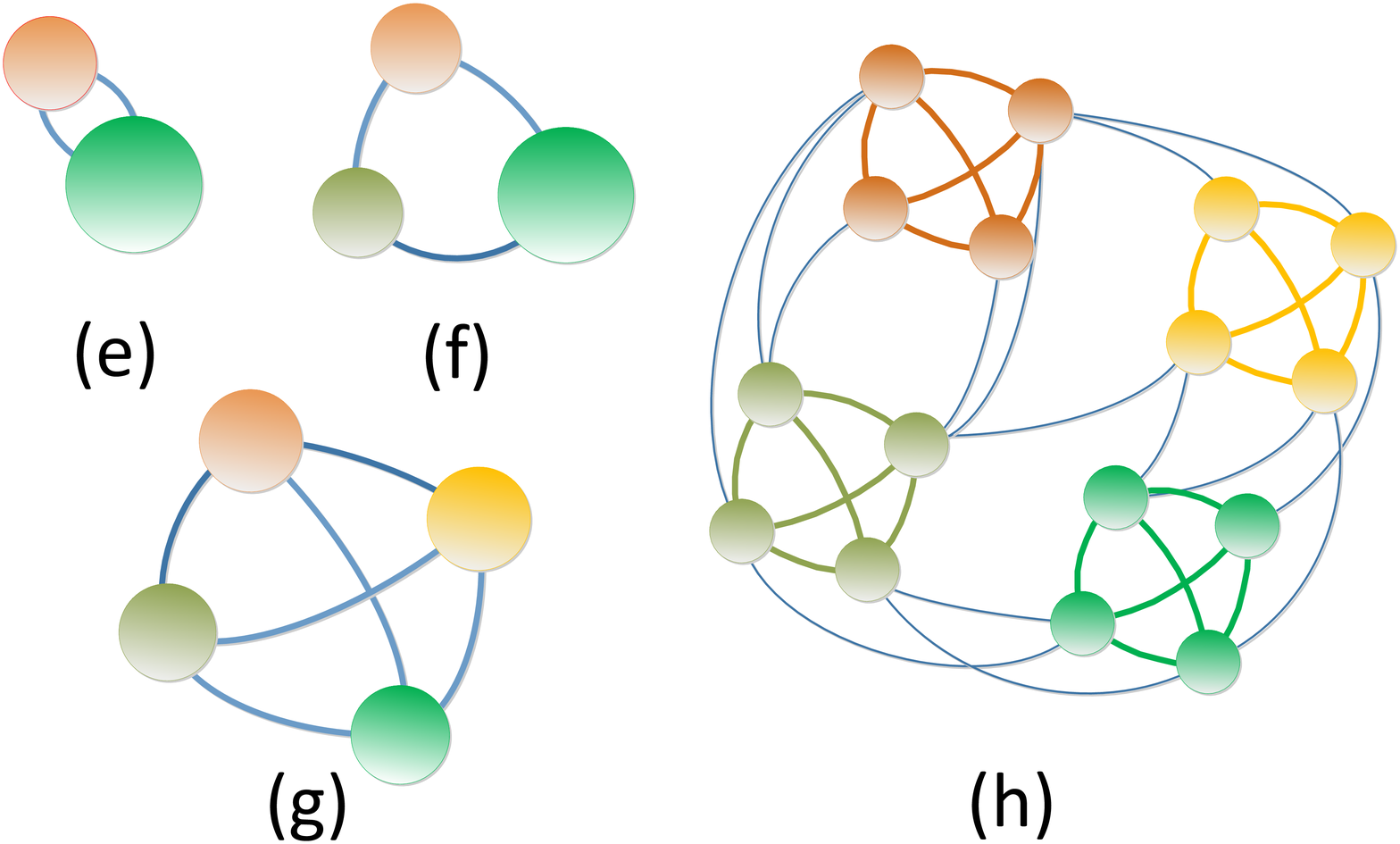}}
\hspace{0.3in} \caption{(Color online) Plots ((a), (b), (c) and (d)) show
the hierarchical tree, and plots ((e), (f), (g) and (h)) describe the cluster
structure of networks implied by the hierarchical tree for $m=2, 3,
4$ and $16$ respectively. It's demonstrated clearly that the
hierarchical structure unfolded by synchronizing process change as
coupling strength decreasing.}\label{fig:hs}
\end{figure}

To show the clustering of nodes more explicitly, we present the
dendrogram plots of agglomerative hierarchical trees and
hierarchical structures implied by the hierarchical trees (see Fig.
\ref{fig:hs}). The dendrogram plots of agglomerative hierarchical
trees for $m=2,3,4$ and $16$ are shown in Fig. \ref{fig:hs} (a),
(b), (c) and (d), respectively, and the corresponding hierarchical
structures of network are described in Fig. \ref{fig:hs} (e), (f),
(g) and (h). When $m=2$, the eigenvector $v_2$ is dominant and
the distances between nodes is to a large extent determined by
it. The network in this case is partitioned into two asymmetric
clusters, and one is much larger than the other one (corresponding
to the distance matrix of $m=2$ in Fig. \ref{fig:dis}). Furthermore,
when $m$ increases from 2 to 4, the four non-overlapping components
at the second level of hierarchical network emerge one by one. At
last, the 16 non-overlapping components at the first level and 4
non-overlapping components at second level are totally unfolded as
the coefficients $\xi_i$ corresponding to eigenvalues from
$v_2$ to $v_{16}$ all diverge, as shown in Fig. \ref{fig:hs}(d) and
(h).

\begin{figure}[!t]
\centering {\includegraphics[width=2.6in]{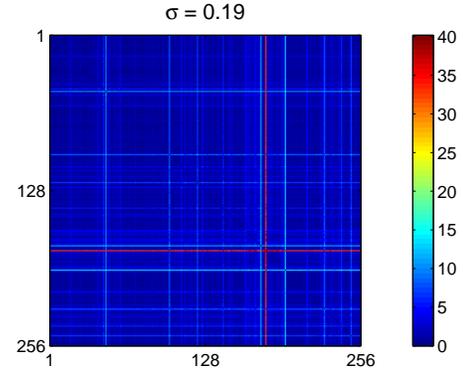}} \caption{
(Color online) Distance matrix with coupling strength $\sigma =
0.19$ at $T=100$. No hierarchical structure
can be seen in this matrix. }\label{fig:du}
\end{figure}

We have demonstrated how the synchronization process of coupling
chaotic oscillators can unravel the hierarchical structure of
network at different levels by tuning the coupling strengths
under the lower boundary of stable region. On the other hand, the
global synchronization of R\"ossler oscillators is also broken if
the coupling strengths exceed the upper boundary of stable region.
In this situation, states vectors are linear combinations of
eigenvectors corresponding to the largest eigenvalues, of which the
elements are almost randomly distributed rather than clustered according
to the substructures of network. Thus, the hierarchical
structure of network cannot be revealed by the distance matrices.
In the simulation, we set the coupling strength in agreement with the the inequality
$\sigma\lambda_m<\alpha_2<\sigma\lambda_{m+1} (m=2,3,...N-1)$. To
eliminate the influence from the eigenvectors corresponding
to small eigenvalues, the coupling strength is set to be larger than
$\alpha_1/\lambda_2\simeq0.17$. For instance, if the coupling
strength is chosen as $\sigma = 0.19$, the smallest value of $m$
that satisfies $\sigma\lambda_m<\alpha_2<\sigma\lambda_{m+1}$ is
$m=229$, which suggests that $\xi_i$ with $i>229$ diverges. The
synchronization process performed on the same hierarchical network
also evolves for time $T=100$, and the distance matrix is also
averaged over a hundred realizations. As shown in Fig. \ref{fig:du},
the visualization of distance matrix shows no evidence of clustering
of nodes. Most values in distances matrix are trivial, because
almost all the states vectors are synchronized to the same state.
Only few nodes are dramatically far away from the others. Therefore,
although the global synchronization is also impossible because
coupling strength exceeds the upper boundary of stable region,
it does not provide useful information to unfold the
hierarchical structure of networks.

\begin{figure}[!t]
\centering {\includegraphics[width=3in]{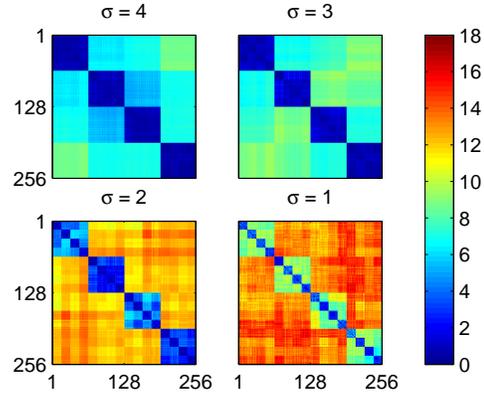}} \caption{
(Color online) Distance matrix with coupling strength $\sigma = 4$,
$3$, $2$ and $1$ at $T=100$.As coupling strength decreasing,
different levels of hierarchical structures are gradually revealed
by synchronizing process}\label{fig:lo}
\end{figure}

We carried out the comprehensive numerical experiments using Lorenz
oscillators to demonstrate the generality of our theoretical
analysis. The state vector and differential equations of Lorenz
oscillators can be written as $\mathbf{x}=[x,y,z]^T$ and
$\mathbf{F}=[10(y-x),x(28-z)-y,xy-\frac{8}{3}z]^T$ respectively. The
units are also coupled by the linear coupling function
$\mathbf{H(x)}=[x,0,0]^T$. The stable region of linear coupling
Lorenz oscillators is $[\alpha_1 \infty]$. Note that the numeric method
gives $\alpha_1=2$, but the accurate value is much lareger in
the numeric simulation of synchronization
process, and we therefore don't set the
coupling strength according to Eq. \ref{coupling}.
Furthermore, because the upper
boundary of stable region of Lorenz oscillators is infinite
($\infty$), we only need to consider the low boundary of the stable region,
and vary the coupling strength $\sigma$ from 4 to 1 for
instance. The synchronization processes are performed on the 14-3
hierarchical network with the same conditions used in
coupling R\"ossler oscillators. The distance matrices of nodes at
$T=100$ are shown in Fig. \ref{fig:lo}. We can see that the
coupling Lorenz oscillators on the hierarchical network beyond global
synchronization region can gradually unfold the substructures and
different hierarchical levels of the network.

In conclusion, we have both theoretically and numerically
investigated the synchronization processes on hierarchical
networks beyond stable region (i.e., the coupling strength is under
the low boundary $\alpha_1/\lambda_2$), which can be used to unfold
the fine substructure such as different levels of hierarchies in an
efficient way. The results show that the nodes belonging to the
same component are tightly clustered according to the
state-vector distances between the corresponding coupling
oscillators. We also find that if the coupling strength exceeds the
upper boundary $\alpha_2/\lambda_N$, the hierarchical structure of
network cannot be effectively unfolded as the trajectories of the
oscillators mix together. Since the nodes belonging to the same
community can be directly identified by their state vectors, our
approach can detect the hierarchical community structure of
complex systems with very low computational complexity, compared to
traditional cross-correlation based or spectral approaches. Our
approach to unravel the hierarchical structure of network
suggests that the dynamics of the interaction contains abundant
information about the topology of the interaction, which can be
utilized to infer the fine structure of complex system.

This work is supported by the National Natural Science Foundation of China under
Grant Nos. 60874090, 60974079, and 61004102. S-MC appreciates the financial support of
the Fundamental Research Funds for the Central Universities (No.WK2100230004). JZ
acknowledges the financial support from the National Natural
Natural Science Foundation of China under grant Nos. 61004104 and 61104143.

\end{document}